\documentclass[twocolumn]{revtex4}
\usepackage{amsmath,lscape,epsfig}

\def\beq{\begin{equation}}
\def\eeq{\end{equation}}
\def\beqa{\begin{eqnarray}}
\def\eeqa{\end{eqnarray}}
\def\ban{\begin{eqnarray*}}
\def\ean{\end{eqnarray*}}
\def\bi{\begin{itemize}}
\def\ei{\end{itemize}}

\begin{document}

\title{Low density instabilities in asymmetric nuclear matter within QMC with
  $\delta$-meson}

\author{Alexandre M. Santos}
\affiliation{Centro de F\'{\i}sica Computacional - Dep. de F\'{\i}sica -
Universidade de Coimbra - P-3004 - 516 - Coimbra - Portugal}

\author{Prafulla K. Panda}
\affiliation{Indian Association for the Cultivation of Science,
Jadavpur, Kolkata-700 032, India}
\affiliation{Centro de F\'{\i}sica Computacional - Dep. de F\'{\i}sica -
Universidade de Coimbra - P-3004 - 516 - Coimbra - Portugal}

\author{Constan\c ca Provid\^encia}
\affiliation{Centro de F\'{\i}sica Computacional - Departamento de F\'{\i}sica -
Universidade de Coimbra - P-3004 - 516 - Coimbra - Portugal}

\begin{abstract}
In the present work we include the isovector-scalar $\delta$-meson in the quark-meson
coupling model (QMC) and study the properties of asymmetric nuclear within QMC
without and with the $\delta$-meson. Recent constraints set by isospin diffusion
on the slope parameter of the nuclear symmetry energy at saturation density are used to adjust the 
model parameters.
The thermodynamical spinodal surfaces are obtained
 and  the instability region at subsaturation densities within  QMC and QMC$\delta$ models are compared with  mean-field relativistic models. The distillation effect in the QMC model is discussed.
\end{abstract}

\maketitle

\vspace{0.50cm}
PACS number(s): {21.65.-f, 21.30.-x, 95.30.Tg}
\vspace{0.50cm}

\section{Introduction}

The instabilities presented by a system are directly related with the possible
phase transitions it can undertake.
At subsaturation densities a
liquid-gas phase transition in nuclear matter is predicted and it is normally tested in nuclear reactions.
The formation of highly excited composed nuclei in equilibrium with a gas of
evaporated particles can be interpreted in the framework of hydrodynamics as
two coexisting phases of nuclear matter, a liquid and a gas phase. During these
reactions, phase transitions may occur depending on the temperature and
densities involved.
The liquid-gas phase transition also plays an
important role in the description of the crust of compact star matter at densities between $0.03$ fm$^{-3}$ and saturation density
($\sim 0.15$ fm$^{-3}$). It essentially consists of neutron rich nuclei immersed in a gas of
neutrons.
It has been shown that this phase transition  leads to an isospin distillation
phenomenon: the isospin content of
each phase is different, most of the gas being composed of neutrons and the
liquid being closer to symmetric matter \cite{S_Xu_00}

In the present paper, we employ the quark-meson coupling model
(QMC)~\cite{guichon,qmc} to investigate the thermodynamical instabilities of
asymmetric nuclear matter. In the QMC model, nuclear
matter is described as a system of non-overlapping MIT bags which interact
through the exchange of scalar and vector mean fields.
An earlier study of ANM within this model has been focused on the effect of isospin
asymmetry and temperature on the equation of state and on the coexistence
surface \cite{pmpk}. We here consider an extension of the model that includes the
 scalar isovector virtual $\delta (a_0(980))$ field \cite{liu}.
Its presence introduces in the isovector channel the structure of relativistic
interactions, where a balance between a scalar (attractive) and a vector
(repulsive) potential exists. The $\delta$- and $\rho$-mesons
give rise to the corresponding attractive and repulsive potentials in
the isovector channel. The introduction of the $\delta$-meson will affect both
 the behaviour of the system at low and high densities. In the last case due to
Lorentz contraction, its contribution is reduced, leading to a harder
equation of state (EOS) at densities larger than $\sim 1.5\, \rho_0$
\cite{cp_07}. At low densities a reduction of the symmetry energy will occur
which will allow for more asymmetric matter.

In  \cite{inst04,thermo06}  the
instabilities in ANM have been investigated  within relativistic mean field hadron models,
both with constant and density-dependent couplings at zero and finite
temperatures. It was shown that the main differences occur at
 large isospin asymmetry and at finite temperature.
In particular it has been shown that the predicted
density at the inner edge of the crust of a compact star, from the crossing of the $\beta$-equilibrium equation of state (EOS), is model dependent \cite{floripa07}.

In the present work we investigate thermodynamical instabilities within the QMC
model with and without the isovector-scalar $\delta$-meson. Although in this model
the isoscalar vector channel described by the $\rho$ meson is included in a
similar way to the non-linear Walecka model (NLWM), the non linearities in the
$\sigma$ and $\delta$-fields arise from the minimization of the bag energy. In
particular, the NLWM used in \cite{liu} does not include non-linearities in the
$\delta$-meson. We may therefore expect a different behavior of asymmetric
matter.

The paper is organized as follows: in section II an extension of the QMC model
to include the $\delta$-meson is discussed, in section III we make a short
review of the calculation of the spinodal surface, in section IV results are
presented and discussed and some conclusions are drawn in the last section.

\section{The quark-meson coupling model}
In what follows we present a review of the QMC model and its generalization
to include the iso-vector-scalar $\delta$-meson.

In the QMC model, the nucleon in nuclear medium is assumed to be a
static spherical MIT bag in which quarks interact with the scalar ($\sigma$, $\delta$)
and vector ($\omega$, $\rho$) fields, and those
are treated as classical fields in the mean field
approximation (MFA) \cite{guichon,qmc}.
The quark field, $\psi_{q_{i}}$, inside the bag then
satisfies the equation of motion:
\begin{eqnarray}
\left[i\,\rlap{/}\partial \right.&-&(m_q^0-g_\sigma^q\, \sigma-g_\delta^q \tau_z 
\delta_3)-g_\omega^q\, \omega\,\gamma^0\nonumber \\
&+&\left. \frac{1}{2} g^q_\rho \tau_z \rho_{03}\gamma^0\right]
\,\psi_{q_{i}}(x)=0\ , \quad  q=u,d
\label{eq-motion}
\end{eqnarray}
where $m_q^0$ is the current quark mass, and $g_\sigma^q$, $g_\delta^q$,
$g_\omega^q$ and $g_\rho^q$ denote the quark-meson coupling constants.
The normalized ground state for a quark in the bag is given
by
\begin{eqnarray}
\psi_{q_{i}}({\bf r}, t) &=& {\cal N}_{q_{i}} \exp 
\left(-i\epsilon_{q_{i}} t/R_i \right) \nonumber \\
&\times& \left(
\begin{array}{c}
  j_{0_{i}}\left(x_{q_{i}} r/R_i\right)\\
i\beta_{q_{i}} \vec{\sigma} \cdot \hat r j_{1_{i}}\left(x_{q_{i}} r/R_i\right)
\end{array}\right)
 \frac{\chi_q}{\sqrt{4\pi}} ~,
\end{eqnarray}
where
\begin{equation}
\epsilon_{q_{i}}=\Omega_{q_{i}}+R_i\left(g_\omega^q\, \omega+
\frac{1}{2} g^q_\rho \tau_z \rho_{03} \right)  ~; ~~~
\beta_{q_{i}}=\sqrt{\frac{\Omega_{q_{i}}-R_i\, m_q^*}{\Omega_{q_{i}}\, +R_i\, m_q^* }}\ ,
\end{equation}
with the normalization factor given by
\begin{equation}
{\cal N}_{q_{i}}^{-2} = 2R_i^3 j_0^2(x_q)\left[\Omega_q(\Omega_q-1)
+ R_N m_q^*/2 \right] \Big/ x_q^2 ~,
\end{equation}
where $\Omega_{q_{i}}\equiv \sqrt{x_{q_{i}}^2+(R_N\, m_q^*)^2}$,
$m_q^*=m_q^0-g_\sigma^q\, \sigma -g_\delta^q\tau_{z}\delta_3$, $R_i$ is the
bag radius of nucleon $i$ and $\chi_q$ is the quark spinor. The bag eigenvalue for nucleon $i$, $x_{q_{i}}$, is determined by the
boundary condition at the bag surface
\begin{equation}
j_{0_{i}}(x_{q_{i}})=\beta_{q_{i}}\, j_{1_{i}}(x_{q_{i}})\ .
\label{bun-con}
\end{equation}
The energy of a static bag describing nucleon $i$ consisting of three quarks in ground state
is expressed as
\begin{equation}
E^{\rm bag}_i=\sum_q n_q \, \frac{\Omega_{q_{i}}}{R_i}-\frac{Z_i}{R_i}
+\frac{4}{3}\,  \pi \, R_i^3\,  B_N\ ,
\label{ebag}
\end{equation}
where $Z_i$ is a parameter which accounts for zero-point motion
of nucleon $i$ and $B_N$ is the bag constant.
The set of parameters used in the present work is given
in Ref. \cite{qmcparm}. The effective mass of a nucleon bag at rest
is taken to be $M_i^*=E_i^{\rm bag}.$
The equilibrium condition for the bag is obtained by
minimizing the effective mass, $M_i^*$ with respect to the bag radius
\begin{equation}
\frac{d\, M_i^*}{d\, R_i^*} = 0,\,\,i=p,n .
\label{balance}
\end{equation}
The total energy density of the nuclear matter reads
\begin{eqnarray}
\varepsilon &=& \frac{1}{2}m_\sigma^2 \sigma^2
+ \frac{1}{2} m_\omega^2 \omega^2_0
+ \frac{1}{2} m_\rho^2 \rho^2_{03}
+ \frac{1}{2} m_\delta^2 \delta_3^2\nonumber\\
&+&\sum_N \frac{1}{\pi^2} \int_0^{k_N} k^2 dk
\left[k^2 + M_N^{* 2}(\sigma,\delta)\right]^{1/2}
\end{eqnarray}
and the free energy density is given by
$$
{\cal F}=\varepsilon-\mu_p\rho_p-\mu_n\rho_n,$$
where the chemical potentials are given by
$$\mu_p=\sqrt{k_p^2+{M^*_p}^2}+g_\omega \rho+\frac{g_\rho}{2} \rho_{03},$$
$$\mu_n=\sqrt{k_p^2+{M^*_n}^2}+g_\omega \rho-\frac{g_\rho}{2} \rho_{03}.$$

The vector mean field $\omega_0$ and $\rho_{03}$ are determined through
\begin{equation}
\omega_0=\frac{g_\omega (\rho_p+\rho_n)}{m_\omega^2},~~~~~~~~~~
\rho_{03}=\frac{g_\rho (\rho_p-\rho_n)}{m_\rho^2},
\end{equation}
where $g_\omega=3 g_\omega^q$ and $g_\rho= g_\rho^q$.
Finally, the mean fields $\sigma_0$ and $\delta_3$ are fixed by
\begin{equation}
\frac{\partial \varepsilon}{\partial \sigma}=0, ~~~~~~~~~~
\frac{\partial \varepsilon}{\partial \delta_{3}}=0.
\label{sig}
\end{equation}

In order to set the model parameters, we start by fixing the free space bag properties. They are obtained by fitting the nucleon mass and enforcing the stability condition for the bag in free space. We consider two sets of free space parameters, taking an equal proton and neutron mass value in a first moment, and then proceeding by considering different proton and neutron masses after.

In the first case, we consider the bare nucleon mass $M=939$ MeV and the bag radius, $R_p=R_n=0.6$ fm. The unknowns $Z_p=Z_n=3.986991$ and $B_N^{1/4}=211.30305$ MeV are then obtained by setting the nucleon bag energies to that (single) bare nucleon mass value.

In the next step, we take the physical nucleon mass values as $M_p=938.272$ MeV and $M_n=939.56533$ MeV and the bag radius for protons as $R_p=0.6$ fm. The unknowns $Z_p=3.98865$, $Z_n=3.98471$, $B_N^{1/4}=211.26209$ MeV and the neutron radius $R_n=0.6002$ are then obtained. Note that for fixed proton bag radius $R_p=0.6$ we observe a decrease on $Z_n$ and on the bag parameter $B_N^{1/4}$ for the nucleons. Next, we fit the quark-meson coupling constants $g_\sigma^q$, $g_\delta^q$, $g_\omega = 3g_\omega^q$ and $g_\rho = g_\rho^q$ for the nucleons, so as to obtain the correct saturation properties of nuclear matter, $E_N \equiv   \epsilon/\rho - M = -15.7$~MeV at
$\rho=\rho_0=0.15$~fm$^{-3}$, $a_{sym}=33.7$ MeV. For the couplings, we have
$g_\sigma^q=5.981$, $g_{\omega}=8.954$. In our first case (when no effective mass difference between $p,n$ is considered), $g_{\rho}=8.615$.

\begin{table*}[t]
  \centering
\caption{Nuclear matter properties of the models used in the present work. All quantities are taken at saturation, except the density $\rho_s$ for which the pressure has a minimum and the incompressibility is zero.}
  \begin{tabular}{lclccccccc}
\hline
    Model & $B/A$ & $~\rho_0$ & $K$ & $M^*/M$ & $\mathcal E_{sym}$ & $L$ 
& $K_{\rm{sym}}$ & $K_{\rm{asy}}$ & $~\rho_s$  \\
          & (MeV)&(fm$^{-3}$)& (MeV) & & (MeV)& (MeV)& (MeV) & (MeV) & (fm$^{-3}$)\\
\hline
NL3~\cite{nl3}                    &16.3 &0.148 & 269 & 0.60 & 37.4  &118.3  & 101   & -608.8& 0.096\\
NL3$\delta$~\cite{alex_plasmon08} &16.3 &0.148 & 270 & 0.60 & 37.4  &153.1  & 427.1 & -491.5& 0.096\\
TW~\cite{tw}                      &16.3 &0.153 & 240 & 0.56 & 32.0  & 55.3  & -125  & -456.8& 0.096\\
QMC                               &15.7 &0.150 & 291 & 0.77 & 33.7  & 93.5  & -10   & -570.8& 0.098\\
QMC$\delta$                       &15.7 &0.150 & 291 & 0.77 & 34.2  &102.1  & 34.8  & -577.6& 0.098\\
BHF~\cite{Isaac_08}               &14.7 &0.182 &176.5& 0.79 & 33.2  & 63.4  & 6.04  & -374.3& 0.119\\
\hline
  \end{tabular}
  \label{tab:properties}
\end{table*}
%
The properties of asymmetric nuclear matter have recently
 been related to both terrestrial data and star properties from Vela pulsar glitches, 
which sets the symmetry energy slope value to $L=88\pm 25$ MeV \cite{xu_08,Xu_arxiv08}.
 We then consider the $\delta$-meson and determine the values of the couplings
 so as to have $L=102.077$ MeV, which sets $g_{\rho N}=12.599$ and $g_\delta^q=12.6$.
In this case, the ($p,n$) mass splitting manifests in the different values 
for the effective masses: $M_p^*=727.718$ MeV and $M_n^*=729.007$ MeV, at saturation.

We take the standard values for the meson masses, namely $m_\sigma=550$ MeV,
$m_\omega=783$ MeV, and $m_\rho=770$ MeV.

\section{Stability Conditions}

The stability conditions for asymmetric nuclear matter, keeping 
constant volume and temperature are obtained from the 
free energy density $\cal F$, imposing that this function is a
convex function of the densities $\rho_p$ and $\rho_n$, i.e. the symmetric
matrix with elements
\begin{equation}
{\cal F}_{ij}=\left(\frac{\partial^2{\cal F}}{\partial \rho_i\partial\rho_j}
\right)_T,
\label{stability}
\end{equation}
is positive \cite{ms,bctl98,mc03}.
This is equivalent to imposing
\begin{equation}\frac{\partial \mu_p}{\partial\rho_p}>0,
\quad
\frac{\partial(\mu_p,\mu_n)}{\partial(\rho_p,\rho_n)}>0, \label{stab2}
\end{equation}
where we have used $\mu_i=\left.\frac{\partial{\cal F}}{\partial \rho_i}
\right|_{T,\rho_{j\ne i}}$.

The two eigenvalues of the stability matrix are given by \cite{ms}
\begin{equation}
\lambda_{\pm}=\frac{1}{2}\left(\mbox{Tr}({\cal F})\pm\sqrt{\mbox{Tr}({\cal
      F})^2-4\mbox{Det}({\cal F})}\right),
\end{equation}
and the eigenvectors $\boldsymbol{\delta\rho_\pm}$ by
$$\frac{\delta\rho^\pm_i}{\delta\rho^\pm_j}=\frac{\lambda_\pm-{\cal F}_{jj}}
{{\cal F}_{ji}}, \quad i,j=p,n.
$$
The largest eigenvalue is always positive whereas the 
other can take on negative values. We are interested in the latter, as it 
defines the spinodal surface,
which is determined by the values of $T,\,  \rho,$ and $y_p$
 for which the smallest eigenvalue  of ${\cal F}_{ij}$ becomes negative. The associated eigenvector defines the instability direction of the system, in isospin space.

It has recently been argued \cite{mc03} that in ANM the spinodal
instabilities
cannot be separately classified as mechanical or chemical instabilities. In
fact, the two conditions that give rise to the instability of the system
are coupled so that the instability appears as an admixture of nucleon density and
concentration fluctuations.
In the following we study the direction of instability and the spinodal for the
different models considered.

\section{Results and discussions}

In the present section we compare the model properties of QMC and QMC$\delta$, respectively with and without the $\delta$-meson, 
the non-linear Walecka model (NLWM) NL3 \cite{nl3}, with and without the $\delta$-meson,
 and the density dependent relativistic hadron model TW \cite{tw}.
We will also refer to the nuclear matter properties obtained  within a microscopic Brueckner-Hartree-Fock (BHF) approximation using the realistic Argonne V18 nucleon-nucleon potential plus a three-body force of Urbana type \cite{Isaac_08}.
\begin{figure}[ht!]
\includegraphics[width=8cm,angle=0]{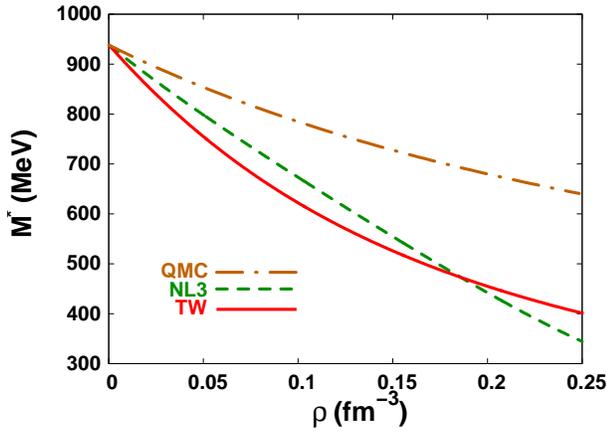}
\caption{(Color online) Effective mass for QMC (brown, dot-dashed), NL3 (green, dashed) and TW (red)
 in symmetric matter.}
\label{fig1}
\end{figure}

\subsection{Model properties}

We will first compare the equilibrium properties of nuclear matter described by the different  models considered. The parameters of these models have been fitted to similar binding energy and saturation density values as seen in Table \ref{tab:properties}. At saturation, the effective mass in QMC is much larger than the corresponding mass in the other models, which is a characteristic of the model \cite{qmc}. In Fig. \ref{fig1}, it is seen that the QMC mass decreases much slower with density. Even the hadronic models we study show quite different behaviour among themselves. NL3 has an almost linear decrease on the mass whereas TW has  much faster drop at low densities, and shows a less dramatic fall as density increases, crossing the curve for NL3 at  $\rho\sim$0.18 fm$^{-3}$. Incompressibility is one of the bulk properties that distinguishes the different models, but it is on the isovector channel that lies the largest distinctions among the different models we use. Although having identical (or barely different) values for their bulk isoscalar  properties, similar models differ considerably on the isovector parameters we discuss next.
\begin{figure}
\begin{tabular}{ll}
\includegraphics[width=8cm,angle=0]{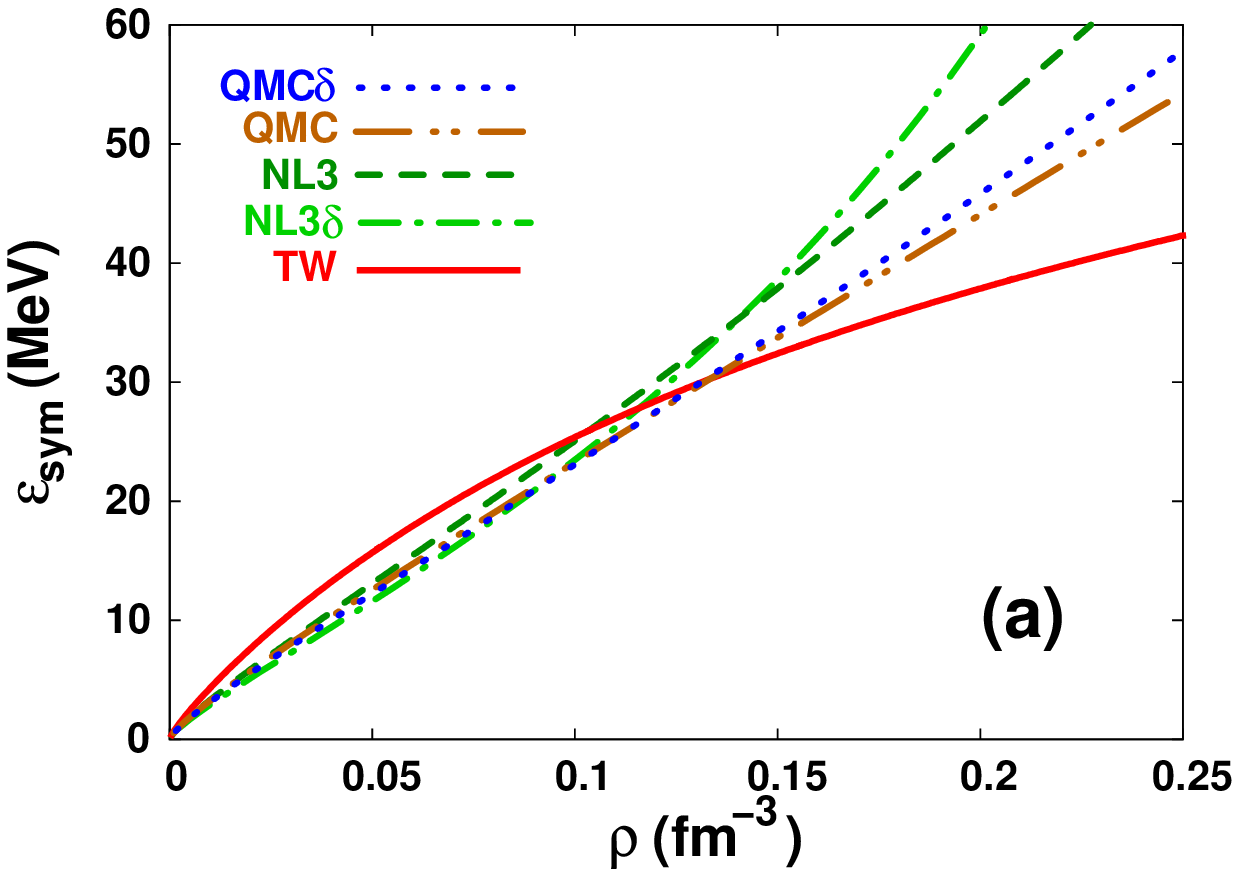}\\
\includegraphics[width=8cm,angle=0]{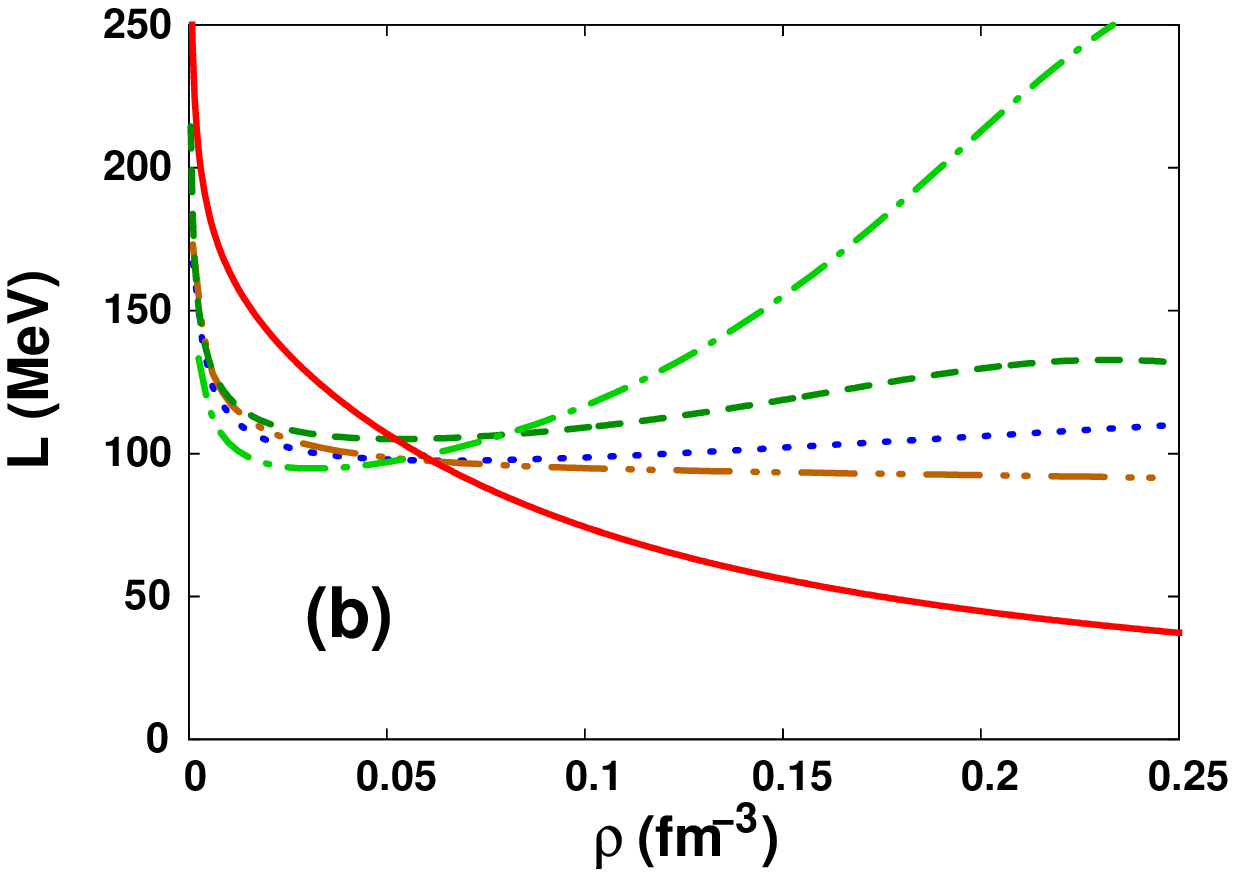}\\
\includegraphics[width=8cm,angle=0]{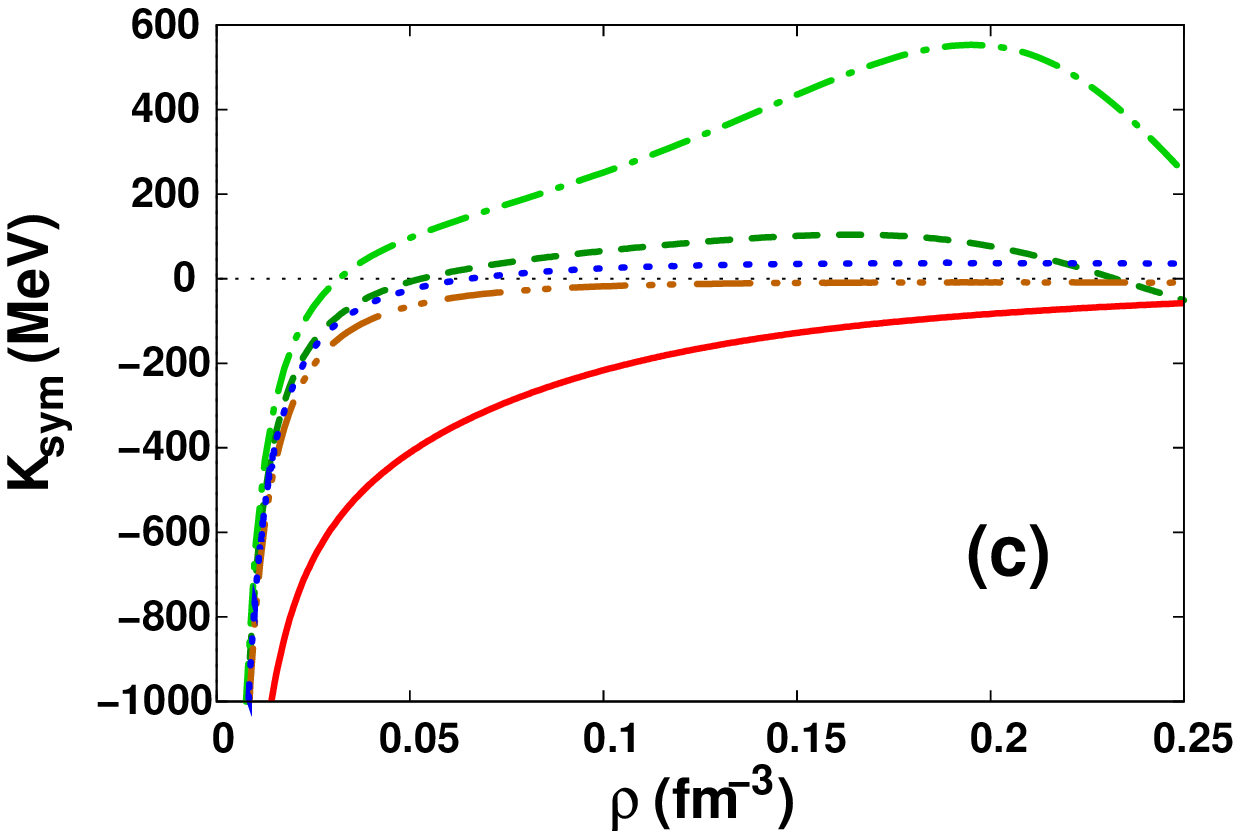}
\end{tabular}
\caption{(Color online) Symmetry energy (a) and its slope parameter 
$L=3\rho_0 {\cal E}^\prime_{sym}$ (b), and $K_{sym}$ (c) in the QMC$\delta$ (blue, dotted), QMC (brown, dot-dot-dashed), NL3 (dark green, dashed), NL3$\delta$ (light green, dot-dashed) and TW (red) models, for symmetric matter.}
\label{fig2}
\end{figure}

We now compare the symmetry energy and its slope and compressibility for all models in this work (Fig. \ref{fig2} and Table \ref{tab:properties}).
The symmetry energy in our relativistic mean field models is given by
\begin{equation}
{\mathcal E}_{sym}=\frac{{k_F}^2}{6{\epsilon_F}^2}+\frac{\rho}{2}\left[\frac{g_\rho^2}{4m_\rho^2}
 - \frac{g_\delta^2}{m_\delta^2}\left(\frac{M_{0}^*}{{\epsilon_F}}\right)^2\right],
\label{esym}
\end{equation}
where $\epsilon_F=\sqrt{P_F^2+M_{0}^{*2}}$ is the Fermi energy of the nucleons, and $M_{0}^*$ is their effective
 mass in symmetric matter.  The NL3 models have the largest  value at saturation of the models considered, 37.4 MeV.

The symmetry energy slope $L(\rho)$ is defined by $L=3\rho_0 \partial{\cal E}_{sym}/\partial\rho.$ 
 The curvature parameter of the symmetry energy $K_{sym}=9 \rho_0^2 \partial^2 {\cal E}_{sym}/\partial\rho^2$ (Fig. \ref{fig2}c)) is also of interest because it distinguishes between different parametrizations. In particular, the quantity  $K_{asy}=K_{sym}-6L$ can be directly extracted from  measurements
 of the isotopic dependence of the  giant monopole resonance (GMR) \cite{bao08}. Recent measurements of the GMR on even-A Sn isotopes give a quite stringent value of $K_{asy}=-550\pm 100$ MeV. According to this value,  the hadronic and QMC models we use here (see Table \ref{tab:properties}) satisfy the above constraint, whereas the BHF results lie slightly below.

The symmetry energy within QMC and QMC$\delta$ shows an extremely linear behaviour with density (Fig. \ref{fig2}a)),
 in comparison with all hadron models shown. This is quite visible from the 
symmetry energy curves, but undoubtedly clear from the slope parameter $L$ (Fig. \ref{fig2}b).
At larger densities the symmetry energy in QMC  is essentially defined by the second term of Eq. (\ref{esym}), proportional to the density, due to the 
small variation of the nucleon effective mass with density. Although still quite hard above  saturation density, the QMC symmetry energy is softer than NL3, but harder than  TW. At subsaturation densities and considering only models without the $\delta$-meson, the QMC symmetry energy takes the smallest values. The introduction of the $\delta$ has the expected effect: at subsaturation densities the symmetry energy is softer but above saturation values it becomes harder due to the saturation of the $\delta$-meson field \cite{liu}. In Fig. \ref{fig2} we show both the NL3$\delta$ and QMC$\delta$ symmetry energies. The effect of the $\delta$-meson on the QMC at subsaturation densities is quite small, much smaller than the effect seen in NL3. It is above the saturation density that the $\delta$-meson has a larger effect in QMC. From the slope of the symmetry energy it is seen that while for QMC the slope decreases slightly with density, for the QMC$\delta$ model it increases slightly, with a value close to 100 MeV. In the bottom figure we also plot $K_{sym}$. The $\delta$ has a very strong effect in the NL3 model. The QMC model is less affected but in both cases the presence of the $\delta$ increases the symmetry incompressibility $K_{sym}$, becoming slightly positive for QMC$\delta$ while it was slightly negative for QMC. The model TW is presenting the smallest values.

\begin{figure}[b]
\includegraphics[width=8cm,angle=0]{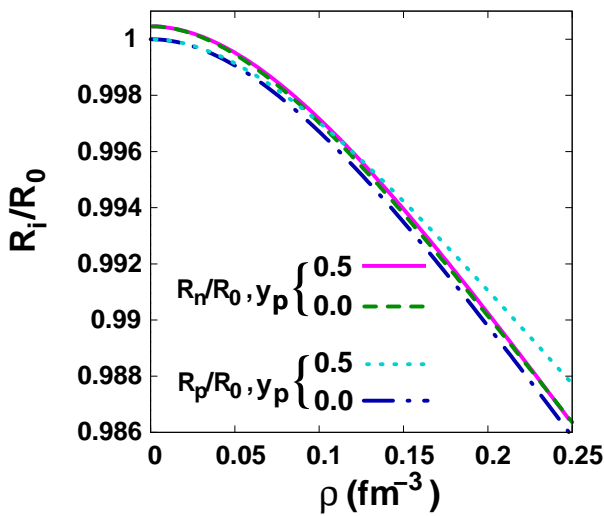}
\caption{(Color online) Radii ratios for neutron and proton, for 
symmetric matter and for $y_p$=0.0 in the QMC$\delta$ model. 
 Here we have taken R=0.6 fm$^{-3}$.}
\label{fig3}
\end{figure}

 Fig. \ref{fig3} shows 
the proton and neutron radii for proton fractions $y_p$=0.5 and 0.0.
For symmetric matter, the neutron bag is larger in this model due to the
mass proton-neutron difference. This result has also been
 reported in \cite{qmc}. Decreasing the proton fraction increases the
proton radius and the  neutron and proton radii cross at a certain value
of the density, isospin dependent:
density:
 $\sim$0.175 fm$^{-3}$
 for $y_p$=0.3, $\sim$0.11 fm$^{-3}$ for $y_p$=0.1 and  $\sim$0.1 fm$^{-3}$ for neutron matter. The bag radius is
sensitive to isospin content by a small amount. It is, however, quite
clear from the neutron matter results ($y_p$=0.0) that the isospin
contents of the nucleons leads to higher radii differences at higher
densities.
Moreover, proton bags become larger than neutrons in medium, as matter
goes denser.
The neutron radius does not change much with isospin and 
for symmetric matter and neutron matter QMC and QMC$\delta$ neutron radii almost overlap.

\begin{figure}[t]
\begin{tabular}{ll}
\includegraphics[width=8cm,angle=0]{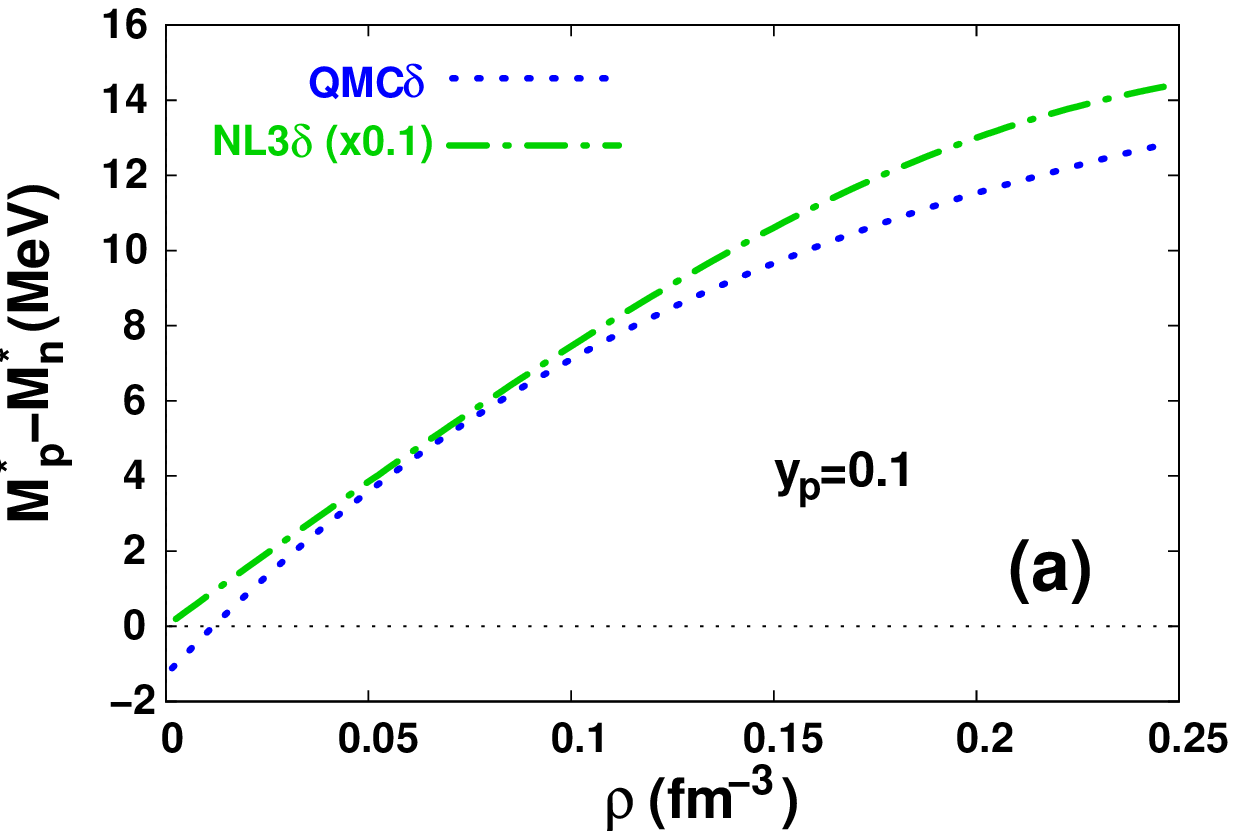}\\
\includegraphics[width=8cm,angle=0]{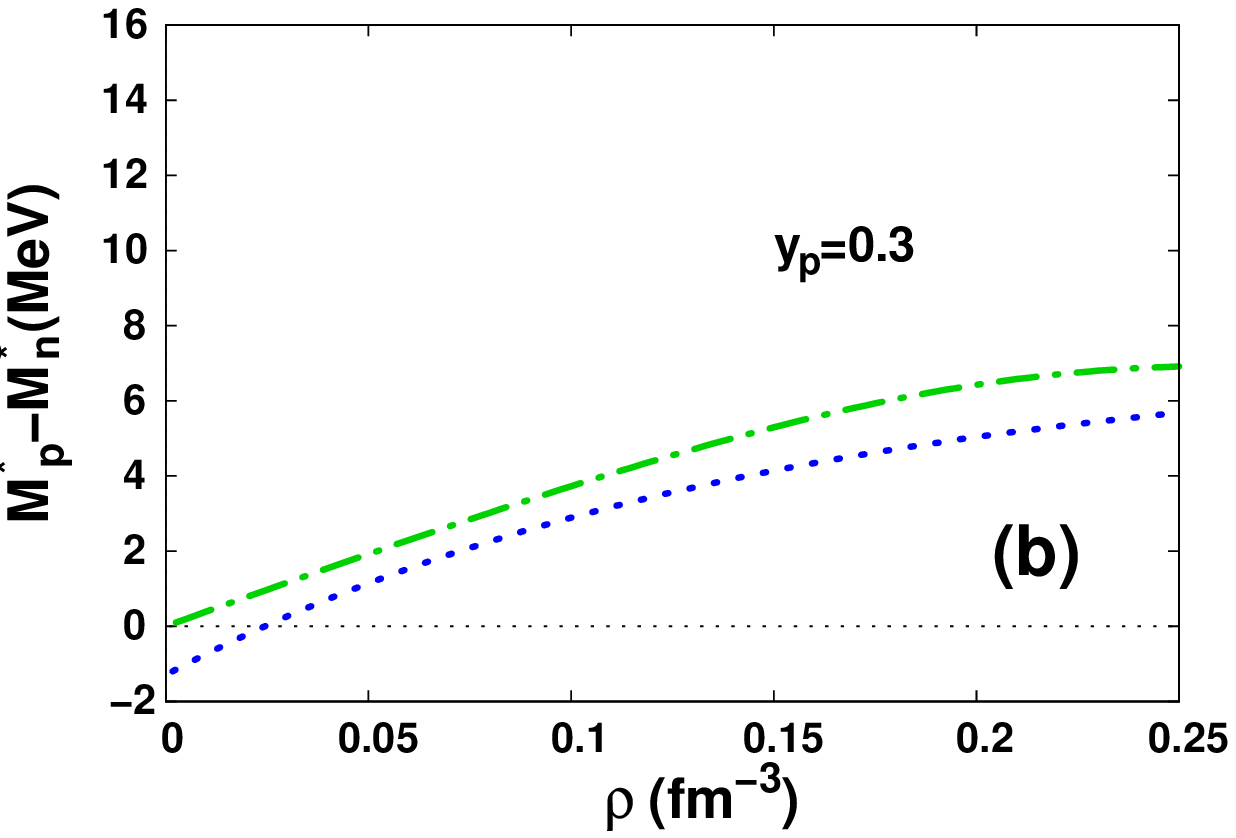}
\end{tabular}
\caption{(Color online) Effective mass differences $M_p-M_n$, for QMC$\delta$ 
(blue, dotted) and NL3$\delta$ (light green, dot-dashed), for different proton fractions: 
$y_p$=0.1 (a) and $y_p$=0.3 (b). Notice that the values 
for NL3$\delta$ have been scaled by a 0.1 factor.}
\label{fig4}
\end{figure}
In Fig. \ref{fig4} we show the effective mass difference $M^*_p-M^*_n$ for
both QMC$\delta$ and NL3$\delta$, for $y_p$=0.3 and 0.1. For NL3, we
show the
 mass difference multiplied by a factor of 0.1 in order  to compare with
QMC$\delta$.
The most striking result is the factor of ten difference between the
$p,n$ mass splitting in NL3 and QMC.
 In both models, this effect increases with baryon density, but it is
worth remarking that
 the effective mass is larger for neutrons than for protons at lower
densities, which is represented by the negative values in the figures.
As referred before, this occurs because the proton and neutron masses
were considered different at zero density. The effect of the
$\delta$-meson is to increase the  $M^*_p-M^*_n$ difference as it occurs
in NL3$\delta$ and other relativistic mean field models and in contrast
to Brueckner-Hartree-Fock calculations \cite{fw06,bhf}
In addition, for the same proton fraction, the crossing of the $p,n$
effective mass curves (equal effective $p,n$ masses) does not occur for
the same densities as
the proton and neutron radius, $R_p$ and $R_n$ (Fig. \ref{fig3}).

\subsection{Instabilities}
\begin{figure}[t]
\includegraphics[width=8.5cm,angle=0]{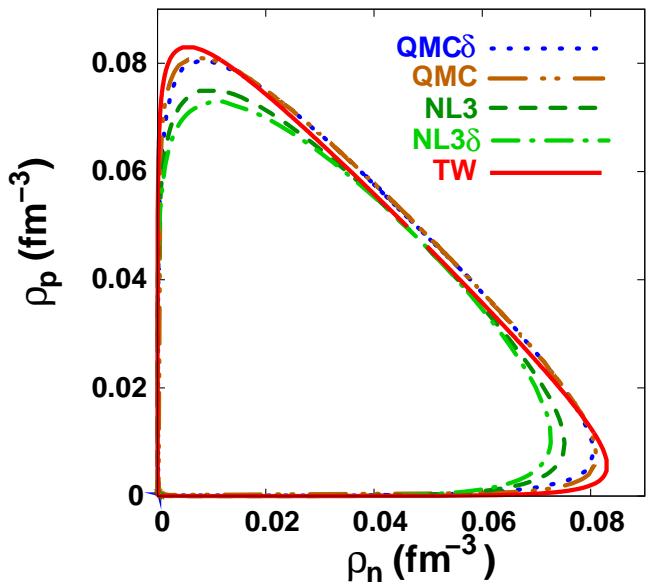}
\caption{(Color online) Spinodal (thermodynamical instability border) for QMC, QMC$\delta$, NL3, NL3$\delta$ and TW.}
\label{fig5}
\end{figure}

In the present subsection we discuss the results for the instability region at subsaturation densities with QMC, QMC$\delta$ and the hadron models we have considered. In Fig. \ref{fig5} we plot the spinodal curves
 for $np$ matter. As referred before, they are defined by the points, for a given temperature, 
density and isospin asymmetry, that make the curvature matrix of the free energy vanish.

Both QMC and QMC$\delta$ present larger
instability regions than NL3, NL3$\delta$ and TW on the isoscalar
direction, $\rho_p=\rho_n$.
This is possibly due to the  $\sigma$ contribution  in the Lagrangian
for QMC (both with and without $\delta$): the fields here attain magnitudes
so as to minimize the bag energy, whereas in hadron models their mean-field
values are determined by solving the relevant set of equations where
nonlinearities
 show explicitly or through density dependent couplings.
The  extension of
the spinodal for $\rho_p=\rho_n$ defines the density $\rho_s$,
corresponding to the density value for which the pressure of symmetric
nuclear matter has a minimum and the incompressibility is zero
\cite{Camille_08}. We have included the values of this density for the
different models in Table \ref{tab:properties}.
Recently \cite{Isaac_08} it was shown that within a
Brueckner-Hartree-Fock calculation the isoscalar extension of
the spinodal was  much larger than NLWM and Skyrme interaction predictions.

\begin{figure}[hb!]
\begin{tabular}{lll}
\includegraphics[width=8cm,angle=0]{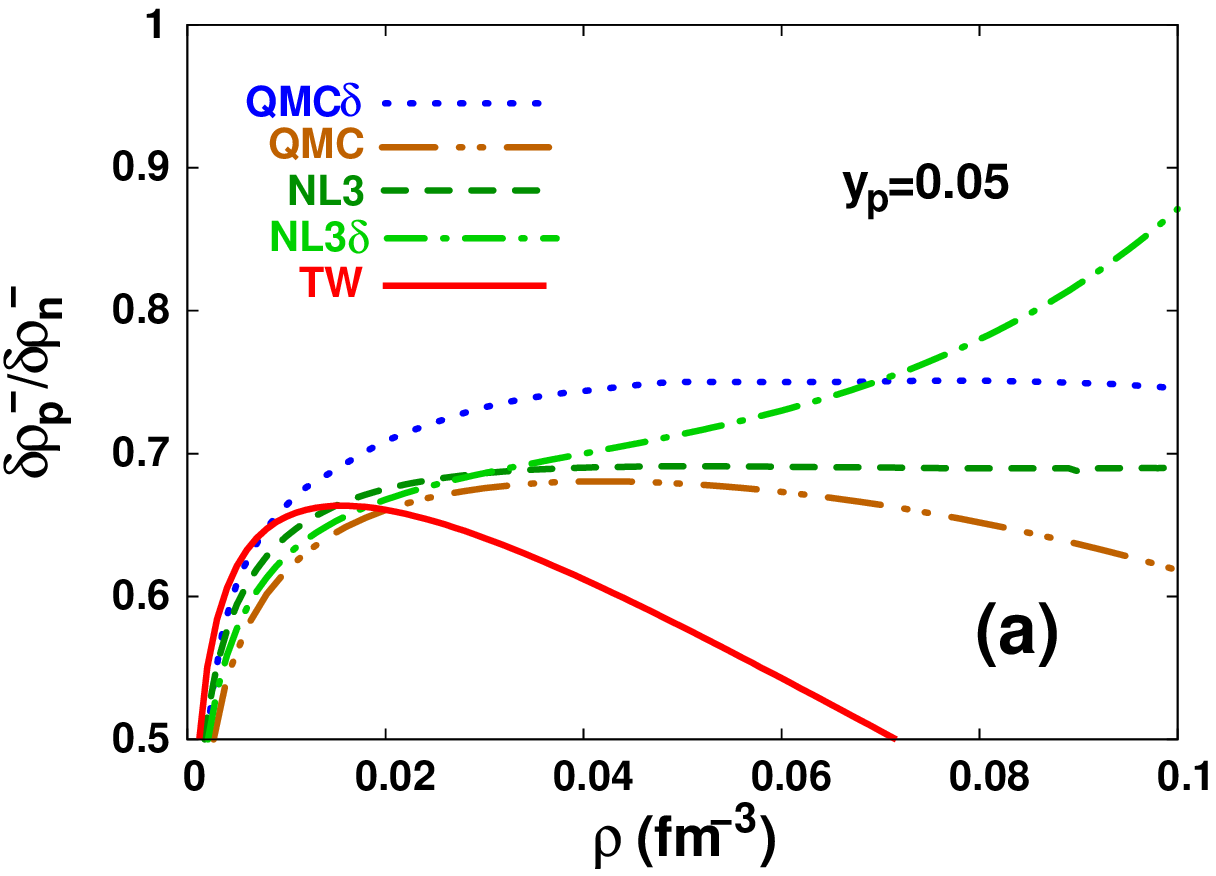}\\
\includegraphics[width=8cm,angle=0]{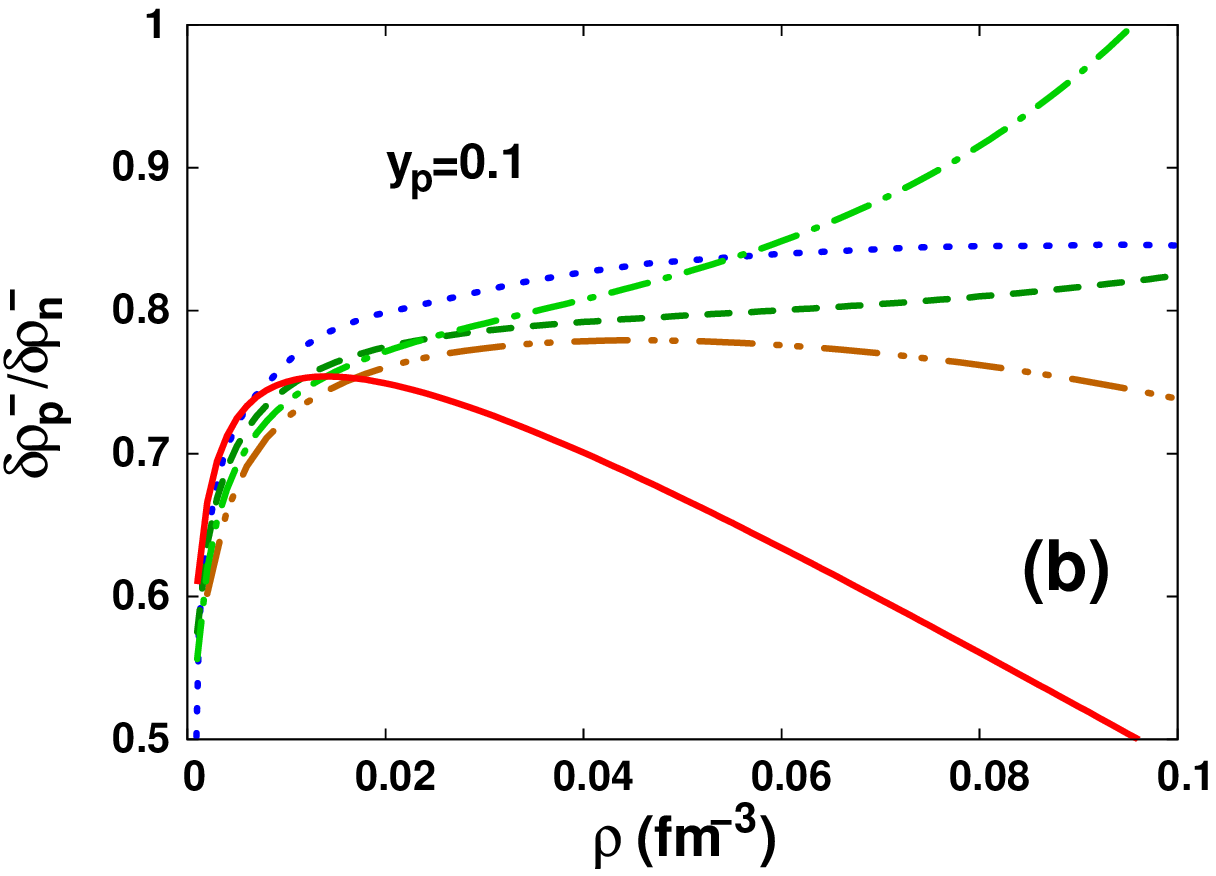}\\
\includegraphics[width=8cm,angle=0]{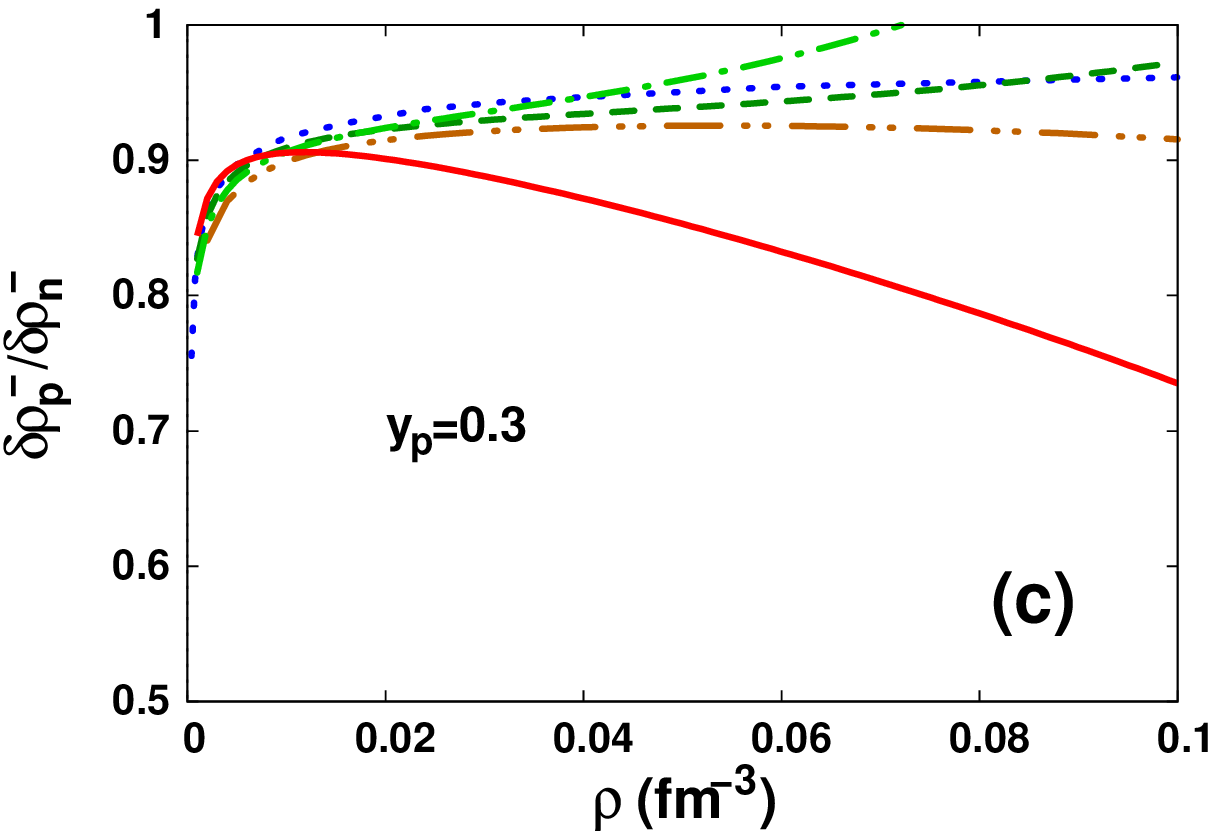}
\end{tabular}
\caption{(Color online) Direction of instability (eigenvector for negative eigenvalue $\lambda_-$) for $y_p$=0.05 (a), 0.1 (b) and 0.3 (c).}
\label{fig6}
\end{figure}
For large isospin asymmetries the presence of
 $\delta$ reduces the instability region both  in QMC and NL3.
 NL3 has a  higher symmetry 
energy than NL3$\delta$ for densities ranging from 0 up to $\sim ~\rho_0$, and the same occurs for QMC although the differences are smaller.
This means that highly asymmetric matter
is less bound in  NL3 and QMC than in NL3$\delta$ and QMC$\delta$.
 A mere inspection of the symmetry energy is not enough to account
 for the differences in the instability region if different models are considered \cite{Camille_08}.
With the introduction of the $\delta$-meson the scalar channel is not affected, and therefore
we are essentially changing the isovector channel. We also notice that at $\rho_p=\rho_n$ the
curvature of the QMC spinodal  is intermediate between NL3 and TW. As shown in
\cite{Camille_08} this curvature is defined by the symmetry energy and its first and second derivatives.

The nuclear liquid-gas coexistence phase is characterized by different isospin contents for each phase, i.e.,
the clusterized regions are more isospin symmetric than the surrounding 
nuclear gas, the so-called isospin distillation \cite{xu_08,chomaz01}. The extension of the distillation effect is model dependent and it has been shown that NL3 and other NLWM parametrizations lead to larger distillation effects than the density dependent hadron models \cite{thermo06,cp_07,Camille_08}. On the other hand,  the distillation effect was also studied within a BHF calculation \cite{Isaac_08} and a smaller distillation effect was generally obtained.
\begin{figure}[b]
\includegraphics[width=8cm,angle=0]{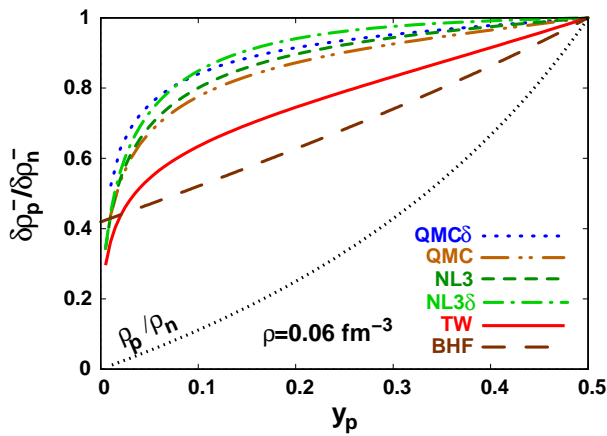}
\caption{(Color online) Proton-neutron density fluctuation ratio versus the isospin
asymmetry for a fixed nuclear density, $\rho$=0.06 fm$^{-3}$. The BHF results were obtained by
Vida\~na and Polls \cite{Isaac_08}. }
\label{fig7}
\end{figure}

In Fig. \ref{fig6} we show the ratio of the proton versus the neutron density fluctuations corresponding to the unstable mode.
This ratio defines  the direction
 of instability of the system. We show the results for different proton fractions
(including rather small values), for the sake of studying 
the effectiveness of the models in restoring the symmetry in the
 liquid phase. 

We  first compare the three models not including the $\delta$-meson: QMC, NL3 and TW. We see that QMC has a behaviour which is intermediate between NL3 and TW. The distillation effect for densities above 0.02 fm$^{-3}$ is larger than the prediction of TW, but for the larger densities it also shows a tendency to decrease, contrary to NL3. The presence of $\delta$-meson  makes the distillation effect more efficient.

In Fig. \ref{fig7} we plot the proton-neutron density fluctuation ratio
as a function of the isospin
asymmetry for a fixed nuclear density, $\rho$=0.06 fm$^{-3}$. We compare all models under
study,  NL3, NL3$\delta$, TW,
QMC and QMC$\delta$, and include also the results  derived from the
BHF approach \cite{Isaac_08} referred above.
All the relativistic models predict larger fluctuation ratios than the corresponding value of
$\rho_p/\rho_n$, dotted line. The behaviour gives rise to a distillation effect, which, as
referred before, is larger for NL3 and smaller for TW, with QMC presenting intermediate values.
The $\delta$-meson stresses the distillation effect, clearly seen both in QMC$\delta$ and NL3$\delta$.

However, except for the very asymmetric matter ($y_p<0.02$),  both  QMC and the other  relativistic
 models predict  fluctuations with larger proton
fractions than BHF. The instability properties within the BFH calculation of  \cite{Isaac_08} 
show also differences for the  spinodal surface: the unstable region is larger, extending to
larger densities both in the isoscalar an isovector directions and  the curvature of the spinodal at $\rho_p=\rho_n$ is much larger than the
one of all the relativistic models considered.  A larger extension of the unstable region is
justified because the saturation density is larger, 0.182 fm$^{-3}$. The shape of the spinodal
itself depends on the density dependence of the symmetry energy and its derivatives. It would be
important  to identify the properties that define the shape of the spinodal for the more  asymmetric matter.

We have studied subsaturation nuclear instabilities for both symmetric
and asymmetric matte within the QMC model, with and without the
inclusion of the $\delta$-meson. In this model the nucleons are
described as non-overlapping bags. We propose a parametrization for QMC$\delta$
with a the symmetry energy slope value $L=102$ MeV within the interval  $L=88\pm 25$ MeV
proposed in \cite{xu_08,Xu_arxiv08} and which was determined from nuclear laboratory data. 
It was interesting to notice that for QMC and QMC$\delta$ the quantity $K_{asy}$ defined in
\cite{xu_08}, and which can be directly extracted from measurements of the isotopic dependence
of the  giant monopole resonances, falls inside the interval predicted by experiments. The BHF calculation
 predicts a value of $K_{asy}$  which is not very far, though lower than the interval obtained from GMR, $K_{asy}=550 \pm 100$ MeV.

A comparison was done with the results
obtained from a NLWM parametrization (NL3), one density dependent
relativistic model (TW), along with BHF with the Argonne V18 potential
calculation.
It was shown that the restoration of isospin symmetry, obtained by a
distillation effect, was more efficient in QMC with respect to TW but
less efficient when compared with NL3. The spinodal surface within QMC
is closer to TW although with a larger curvature at $\rho_p=\rho_n$ and
a slightly smaller instability extension at larger asymmetries, while
the inclusion of the $\delta$-meson (QMC$\delta$) shrinks the asymmetric
parts of the instability envelope. The BHF results of \cite{Isaac_08}, although with similar
general properties, differ in the extension and shape of the spinodal and in the amount of
distillation predicted. 

A study  of the QMC instability properties at  finite temperatures is now under preparation. It
is also important to identify how the density dependence of the symmetry energy determines the
shape of the spinodal for large isospin asymmetries. 
%
\section{ACKNOWLEDGMENTS}
This work was partially supported by FEDER and FCT (Portugal) under the
projects POCI/FP/81923/2007, CERN/FP/83505/2008 and SFRH/BPD/29057/2006. The authors acknowledge I. Vida\~na for kindly providing us with some data for comparison in this work.

\end{document}